# Lateral electron tunneling through single self-assembled InAs quantum dots coupled to superconducting nanogap electrodes

K. Shibata[1,*], C. Buizert[2], A. Oiwa[2], K. Hirakawa[1], and S. Tarucha[2,3]

[1] IIS and INQIE, University of Tokyo, 4-6-1 Komaba, Meguro-ku, Tokyo 153-8505, Japan
[2] Department of Applied Physics and INQIE, University of Tokyo, 7-3-1 Hongo, Bunkyo-ku, 113-8656, Japan
[3] ICORP Spin Information Project, JST

We have fabricated superconductor-quantum dot-superconductor (SC-QD-SC) junctions by using SC aluminum electrodes with narrow gaps laterally contacting a single self-assembled InAs QD. The fabricated junctions exhibited clear Coulomb staircases and Coulomb oscillations at 40 mK. Furthermore, clear suppression in conductance was observed for the source-drain voltage $|V_{SD}| < 2\Delta/e$, where $\Delta$ is the SC energy gap of Al. The absence of Josephson current that flows through QDs is due to the strong Coulomb interaction and non-negligible thermal fluctuation in our measurement system.

Quantum dots (QDs) are often called "artificial atoms" and show varieties of atom-like physics [1]. The physics and applications of single QDs have attracted much attention in the context of their application as quantum information devices. To date, the optical properties of single self-assembled InAs QD have been extensively studied; it has been revealed that InAs QDs have excellent optical properties, large orbital quantization energies, strong carrier-carrier interactions, and spin-related physics, etc. These make InAs self-assembled QD attractive for applications to photonic devices and quantum information processing. However, transport properties of single self-assembled InAs QDs have been much less studied so far [2-5].

It is known that, even when metals are deposited directly on InAs surfaces, the metal Fermi level is often pinned inside the conduction band of InAs [6]. Utilizing this property, it is possible to realize strong coupling between the localized electron wavefunctions in InAs QDs and the electrons in metallic electrodes. Indeed, it was recently reported that self-assembled InAs QDs directly probed by nanogap Au electrodes operate as single electron transistors (SETs) [7], exhibiting clear shell filling [8] and the Kondo effect [9,10]. Furthermore, by replacing nonmagnetic Au electrodes with ferromagnetic metals, spin transport through single self-assembled InAs QD has also been observed [11,12]. These results indicate that InAs QDs are very compatible with metallic electrodes and well suited for realizing additional functionalities by choosing appropriate contacting metal species. By using superconducting (SC) materials as nanogap electrodes, further functionalities such as the gate control of Josephson current can be added to InAs QD SETs. Electron transport through semiconductor nanostructures coupled with superconducting electrodes has been intensively investigated using two dimensional electron gas [13], one dimensional InAs quantum wires [14], and carbon nanotubes [15-18]. These systems are also very good candidates for studying novel transport phenomena, e.g., superconducting proximity effects [13-17], multiple Andreev reflection (MAR) in quantum nanostructures [16,18,19], and competition between superconductivity and magnetism [17,20]. These subjects are worth studying in further details by using self-assembled InAs QDs, which have distinct characteristics mentioned above.

In this work, we have fabricated SC-QD-SC junctions by using aluminum nanogap electrodes laterally contacting a single self-assembled InAs QD. The fabricated junctions exhibited clear Coulomb blockade effects. Furthermore, clear suppression in conductance was observed around $V_{SD} = 0$ V for a voltage range of $4\Delta/e$ at $T = 40$ mK, where $\Delta$ is the SC energy gap of Al.

Self-assembled InAs QDs were grown by molecular beam epitaxy on a (100)-oriented GaAs substrate. After successively growing a 100 nm-thick $Al_{0.3}Ga_{0.7}As$ barrier layer and a 200 nm-thick undoped GaAs buffer layer, self-assembled InAs QDs were grown at 500 °C. The InAs coverage of ~4ML was used to obtain a mixed phase of large and small dots. The larger QDs are more easily contacted, increasing the yield of the fabrication process, and also provide low tunneling resistance [7]. A degenerately Si-doped layer 300nm below the surface was used as a backgate electrode.

The pattern of the electrodes was defined by electron beam lithography in PMMA resist. To avoid electrical contacts through highly resistive natural oxides on the Al surfaces, we first deposited thin Ti(10nm)/Au(30nm) contact terminals and subsequently nanogap Al electrodes (a 5 nm-thick Ti layer and a 100 nm-thick Al layer) were deposited on top of these terminals in an electron beam evaporator. The evaporation was preceded by a 6 second wet etch in buffered hydrofluoric acid to remove any oxide on the QDs to make the contact more transparent. The probability of a single InAs dot bridging the nanogap is approximately 5%. To improve this yield, we have fabricated a series of initially unconnected nanogaps. Figure 1(b) shows a scanning electron microscope (SEM) image of an array of junctions. The inset shows a nanogap containing a single InAs dot bridging the two electrodes. The granular nature of the Al makes the detection of QDs more difficult, as compared with metals that are deposited in smooth films, such as gold [7-10]. SEM observation reveals which gaps contain InAs QDs. Finally, Ti(20nm)/Au(150nm) bridge electrodes which connect successful nanogap electrodes and bonding pads as shown in Fig. 1(a) were deposited on top of the gold terminals, thus creating an oxide-free electrical contact. We adopted an electrode pattern for four-terminal measurement. In these configurations schematically shown in Fig. 1(c), low contact resistance of ~20 Ω was achieved between gold terminals and Al electrodes, and gold terminal and gold bridge electrodes, respectively. These junctions with selected nanogaps show reasonable conductance with a yield of more than 25%.

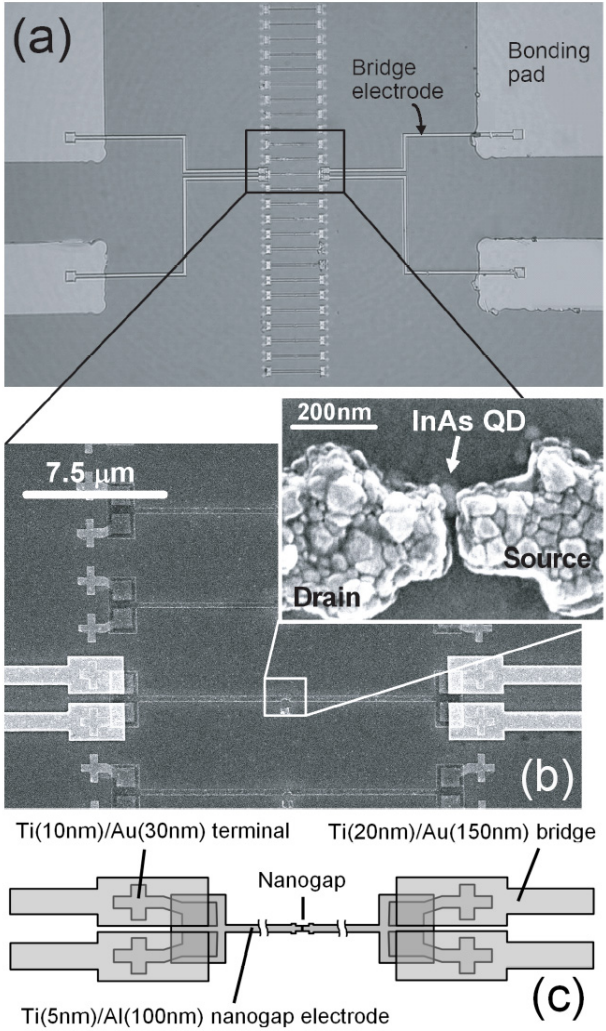

FIG. 1: (a) Optical micrograph of the gold bridge electrodes that connect the contact terminals of the selected nanogap electrodes with the bonding pads. (b) A blowup of an SEM image for the area of terminals and Al electrodes. The inset shows an SEM image of an Al nanogap containing a single InAs QD. (c) Schematic details of the fabricated SC-QD-SC junction.

A device fabricated in the aforementioned fashion was placed in a dilution refrigerator with a base temperature of 40 mK, which was fitted with copper powder filters and second order RC filters (10 kHz cut-off) to prevent instrumentation noise from heating the leads. Figure 2 shows current-voltage (*I-V*) curves of a device with a room-temperature junction resistance of 300 kΩ measured for various backgate voltages. The junction exhibits suppressed conductance near zero bias voltage over the wide voltage range ($V_{SD}$ ~ 10 mV) and the voltage width of the suppressed conductance region is modulated by changing $V_G$. Step-like *I-V* structures, so-called Coulomb staircase, are also seen. These results clearly indicate that this junction operates as a SET. However, from a closer look at Fig. 2, it is noticed that the current suppressed regions in the *I-V* curves shown in Fig. 2 exhibit small but finite slopes in the Coulomb blockade regime. This finite slope is attributed to the co-tunneling process [21]. The signature of the superconductivity in Al electrodes, which will be described later, is not visible at this plot scale.

Figure 3(a) shows a part of the Coulomb stability diagram obtained by taking the differential conductance $dI/dV_{SD}$ as a function of $V_{SD}$ and $V_G$. Provided that we have junction transparency down to the last electron, we obtain an occupancy $N = 10$ for the leftmost diamond from counting the Coulomb oscillations shown in the inset of Fig. 2. We find a charging energy $U_C$ ~ 3 meV in the same order as the level splitting $\delta E$. The electrostatic lever arm factor for the backgate voltage is ~0.055 meV/mV.

As mentioned above, the device is relatively transparent in the chosen range of $V_G$, yielding a detectable dot conductance in the Coulomb blockade regime due to elastic co-tunneling processes. When the bias is reduced below $|V_{SD}| < 2\Delta/e$, this mechanism is suppressed due to the absence of quasiparticle states in the density of the states (DOS) in the SC leads. This shows up in Fig 3(a) as a dark strip around $V_{SD} = 0$. Figure 3(b) shows a more detailed scan of $dI/dV_{SD}$ in the Coulomb blockade region as a function of $V_{SD}$. The various curves correspond to different values of the externally applied magnetic field, from 0 T (lower) to 100 mT (upper) in steps of 25 mT. The magnetic field was applied perpendicular to the substrate. At zero field we observe maxima in the conductance around $V_{SD} = \pm 0.28$ mV. These points correspond to the onset of direct quasiparticle tunneling, as depicted schematically in Fig. 3(c), and its height reflects the singularities in the SC DOS of the leads. From the peak separation of $4\Delta$, we can estimate a SC gap energy of $\Delta = 140$ μeV at zero field, or a critical temperature of $T_c = 0.92$ K via the BCS relation $\Delta = 1.76 \, k_B T_c$ [22]. In a simultaneously deposited Al test strip we could observe a dissipationless supercurrent up to 0.9 K (not shown), which is in good agreement with $\Delta$ found from the tunneling characteristics. The critical magnetic field $B_c$ of aluminum thin film is known to be strongly dependent on film thickness. At 100 mT the signatures of superconductivity have disappeared. This relatively high $B_c$ is consistent with the previously reported value [23]. We have measured several samples and confirmed that they also exhibit similar behaviors. In contrast to other reported studies on nanostructures with SC leads [16,18,24], no subgap structures due to MAR are observed, because they are suppressed by the strong on-site Coulomb interactions between the carriers [24, 25]; i. e., $U_c$ is over 20 times larger than $\Delta$.

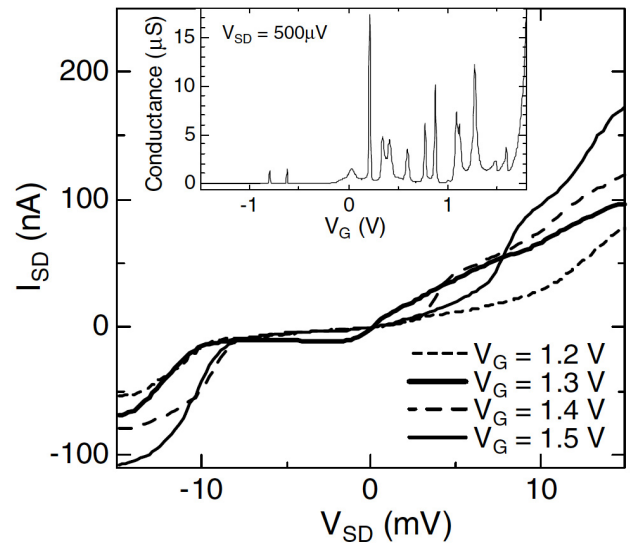

FIG. 2 *I–V* curves measured at 40 mK for a junction with a single InAs QD coupled to Al electrodes. Four *I–V* curves taken at different backgate voltages $V_G$ are shown. The inset shows linear conductance spectrum taken by applying $V_{SD}$ = 500 μV as function of $V_G$.

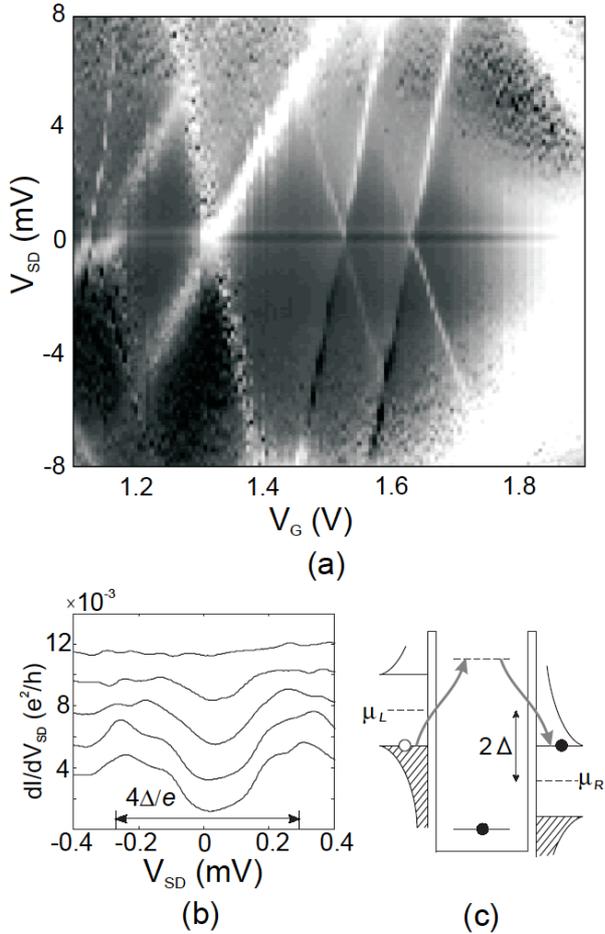

FIG. 3 Signature of superconducting leads in the transport properties of the QD. (a) Gray-scale plot of the differential conductance $dI/dV_{SD}$ through the dot, showing the Coulomb stability diagram. Around zero bias, the elastic co-tunneling is suppressed by the superconducting gap in the DOS of the superconducting Al leads. (b) Detailed scan of the gap structure in the Coulomb blockade regime as a function of magnetic field. The curves correspond to 0, 25, 50, 75 and 100 mT from the bottom to the top. Curves are offset for clarity. (c) Schematic of the co-tunneling process between the superconducting leads. For $|V_{SD}| = 2\Delta/e$ we find a maximum in the differential conductance, reflecting the singularities in the DOS spectrum.

By applying a current-bias across 2 terminals, we can probe the voltage drop over the remaining two to detect if there is any dissipationless flow due to the Josephson effect. Due to the high $U_c \gg \Delta$, the Cooper pairs of charge $2e$ cannot tunnel directly and, consequently, the electrons have to be transferred one by one. This can yield a finite critical current $I_c$ provided that the electrons tunnel in a coherent fashion [26]. In the cited reference, $I_c$ was in the order of a few hundred pA, which corresponds to a Josephson energy $E_J = \hbar I_c/2e$ around 10 mK. This is far below the noise temperature of our setup, which was in the order of 100 mK. The SC phase correlations across the junction are washed out by thermal fluctuations and we cannot observe dissipationless branch in the I-V characteristics. Future attempts will focus on both reducing the effective electron temperature by improving the setup filtering and also increasing $I_c$. The latter can be done by 1) making the junction more transparent for high transport coherence, 2) reducing $U_c$ by selecting larger dots and reducing the nanogap separations, and 3) using different SC materials that have larger gap energies $\Delta$.

In summary, we have fabricated SC-QD-SC junctions by using aluminum nanogap electrodes coupled with a single self-assembled InAs QD. These junctions exhibited single electron tunneling behaviors at $T = 40$ mK. Furthermore, clear suppression in conductance was observed around $V_{SD} = 0$ V for a voltage range of $4\Delta/e$, which reflects the superconductivity in the Al leads. The absence of the Josephson current that flows through QDs is due to the strong Coulomb interaction $U_c \gg \Delta$ and nonnegligible thermal fluctuation in our system.

We thank H. Sakaki, Y. Arakawa, and T. Machida for fruitful discussions. This work was partly supported by the Grant-in-Aid from the Japan Society for Promotion of Science (No. 18201027) and the Specially Coordinated Fund from MEXT (NanoQuine).